\documentclass{aastex62}

\def \kms{\mbox{km\,s$^{-1}$}}

\def \araa{ARA\&A}
\def \mnras{MNRAS}

\def \apjs{ApJS}
\def \apj{ApJ}
\def \apjl{ApJL}
\def \pasa{PASA}
\def \aap{A\&A}

\newcommand{\um}{$\mu$m}

\newcommand{\Msun}{M$_{\odot}$}

\newcommand{\pcc}{${\rm cm}^{-3}$}
\newcommand{\hii}{\mbox{$\mathrm{H\,{\scriptstyle {II}}}$}}

\newcommand{\nthpnt}{N$_{2}$H$^{+}$}
\newcommand{\nthp}{N$_{2}$H$^{+}$\,(1--0)}
\newcommand{\hcopnt}{HCO$^{+}$}
\newcommand{\hcop}{HCO$^{+}$\,(1--0)}

\newcommand{\htcop}{H$^{13}$CO$^{+}$\,(1--0)}

\newcommand{\hnc}{HNC\,(1--0)}

\newcommand{\Pebble}{G337.342$-$0.119}

\received{25 January 2018}

\revised{ }

\accepted{ }

\published{ }

\submitjournal{Astrophysical Journal}
\shorttitle{The Pebble}
\shortauthors{Jackson et al.}

\begin{document}

\author{James M. Jackson}\affiliation{School of Mathematical and Physical Sciences, University of Newcastle, University Drive, Callaghan NSW 2308, Australia}
\affiliation{SOFIA Science Center, Universities Space Research Association, NASA Ames Research Center, Moffett Field, CA, 94035, USA}
\author{Yanett Contreras}\affiliation{Leiden Observatory, Leiden University, P.O. Box 9513, NL-2300 RA Leiden, the Netherlands}
\author{Jill M. Rathborne}\affiliation{CSIRO Astronomy and Space Science, P.O. Box 76, Epping NSW 1710  Australia}
\author{J. Scott Whitaker}\affiliation{Physics Department, Boston University, 590 Commonwealth Ave., Boston, MA, 02215 USA}
\author{Andr{\'e}s Guzm{\'a}n}\affiliation{Departamento de Astronom{\'i}a, Universidad de Chile, Camino el Observatorio 1515, Las Condes, Santiago, Chile}
\affiliation{National Astronomical Observatory of Japan, National Institute of Natural Sciences, 2-21-1 Osawa, Mitaka, Tokyo 181-8588, Japan}
\author{Ian W. Stephens}\affiliation{Radio and Geoastronomy Division, Harvard Smithsonian Center for Astrophysics, MS-42, Cambridge, MA, 02138, USA}
\author{Patricio Sanhueza}\affiliation{National Astronomical Observatory of Japan, National Institute of Natural Sciences, 2-21-1 Osawa, Mitaka, Tokyo 181-8588, Japan}
\author{Steven Longmore}\affiliation{Astrophysics Research Institute, Liverpool John Moores University, 146 Brownlow Hill, Liverpool L3 5RF, UK}
\author{Qizhou Zhang}\affiliation{Harvard-Smithsonian Center for Astrophysics, 60 Garden St., Cambridge MA 02138 USA}
\author{David Allingham}\affiliation{School of Mathematical and Physical Sciences, University of Newcastle, University Drive, Callaghan NSW 2308, Australia}

\correspondingauthor{James M. Jackson}
\email{jjackson@usra.edu}

\title{\Pebble\  (the ``Pebble''): A Cold, Dense, High-Mass Molecular Cloud with Unusually Large  Linewidths and a Candidate High-Mass Star Cluster Progenitor}

\begin{abstract}
Exactly how high-mass star clusters form, especially the young massive clusters (YMCs:  age $<100$ Myr; mass $>10^4$\,\Msun), remains an open problem,
largely because they are so rare that examples of their cold, dense, molecuar progenitors continue to be elusive.
The molecular cloud \Pebble, the ``Pebble,'' is a candidate for such a cold progenitor.  Although \Pebble\, was originally identified as four separate ATLASGAL clumps,
the similarity in their molecular line velocities and linewidths in the MALT90 dataset demonstrate that these four clumps are in fact one single, coherent cloud.
This cloud is unique in the MALT90 survey for its combination of both cold temperatures ($T_{dust} \sim 14$ K) and large linewidths $(\Delta V \sim 10~\kms)$.
The near/far kinematic distance ambiguity is difficult to resolve for \Pebble.  At the near kinematic distance (4.7 kpc), the mass is 5,000 \Msun\, and the size is $7\times2$ pc.
At the far kinematic distance (11 kpc), the mass is 27,000 \Msun\, and the size is $15 \times 4$ pc. The unusually large linewidths of \Pebble\,are difficult to reconcile with a gravitationally bound system in equilibrium.  If our current understanding of the Galaxy's Long Bar is approximately correct,
\Pebble\, cannot be located at its end. Rather, it is associated with a large star-forming complex that contains multiple clumps with large linewidths.  If \Pebble\, is a prototypical cold progenitor for a high-mass cluster, its properties may indicate that the onset of high-mass star cluster formation is dominated by extreme turbulence. 

\end{abstract}


\section{Introduction}

Because stars primarily form within clusters, understanding star formation requires a thorough understanding of cluster formation.
Clusters form with a large but continuuous spread in their masses (cf., \citealt{Bressert10}).
The most extreme examples of current cluster formation are the Young Massive Clusters (YMCs: age$<$100 My and  stellar mass$>\sim10^4$\,\Msun), 
such as the Arches, NGC 3603, Westerlund  1 (see review by \citealt{PotegiesZwart10}).  
The deeply embedded luminous young clusters associated with SgrB2 \citep{ZhaoWright2011, Ginsberg18}, W49A \citep{Saral2015}, W43, and W51A \citep{Saral2017} are almost certainly YMCs in their early stages.
The masses of YMCs fall between those of typical open star clusters and the much more massive 
globular clusters.
Determining the initial conditions of the cold, dense, high-mass, molecular clouds that will eventually produce high-mass clusters, and especially the YMCs,
remains an open question,
largely because these cold, high-mass cluster precursors are extremely rare.
If such precursors can be found, their structure and physical conditions will provide important clues about the cluster formation process, their internal
turbulent structure, the stellar initial mass function, and the highly centrally peaked stellar spatial distribution observed in YMCs   (cf., \citealt{Walker16}). 

Perhaps the best-known example of a candidate YMC molecular precursor is G0.253+0.016 (the ``Brick'': M$\sim$10$^5$\,\Msun\, R$\sim$3 pc, T$_{dust}\sim$ 20 K; 
\citealt{Longmore12,  Rathborne14a, Rathborne14b, Rathborne15}).  The Brick's log-normal column density distribution function matches theoretical predictions
for a clump whose structure is dominated by turbulence \citep{Rathborne14b}.  Moreover, the small power-law tail at high column densities suggests that gravitational
collapse occurs at a specific column density threshold at which self-gravity can overcome turbulent support.  One of the Brick's most intriguing properties is its
combination of low dust temperatures ($\sim 20$ K) and extremely large linewidths ($\Delta V \sim 30$ \kms). This extreme turbulence may arise from its location in the Galaxy's Central Molecular Zone. To further test the agreement between theoretical predictions 
of the structure of dense molecular cluster precursors with observations, it is important to discover new candidates for cold progenitors to high-mass star clusters, and preferably outside of the unusual environment of the Central Molecular Zone.

Over the past several years, thousands of dense molecular clumps have been discovered and characterized by millimeter and submillimeter continuum Galactic surveys, such as the
1.1 mm Bolocam Galactic Plane Survey (BGPS; \citealt{Aguirre11}), the 870 \um\, APEX Telescope Large Area Survey of the Galaxy (ATLASGAL; \citealt{Schuller09}), and the 70 to
500 \um\, {\it Herschel} Infrared Galactic Plane Survey (HiGAL; \citealt{Molinari10}), yet high-mass cluster precursors in the earliest, cold, “pre-stellar” phases remain particularly elusive.
Indeed, \citet{Ginsburg14} conclude that no such cold precursors with mass $M >10^4$ \Msun\, and radius $r<2.5$ pc exist in the first Galactic quadrant and outside of the Galaxy's central kpc.

We have extended the search for cold precursors to high-mass star clusters to the fourth Galactic quadrant.  To search for such precursors, we have analyzed  870\,\um\, ATLASGAL  and 70 to 500 \um\,  HiGAL images to estimate dust temperatures and gas column densities \citep{Guzman15}.  Combined with kinematic distance estimates \citep{Whitaker17} from the Millimetre Astronomy Legacy Team 90 
GHz Survey (MALT90; \citealt{Foster11, Foster13, Jackson13}) molecular line velocity data \citep{Rathborne16}, 
the dust continuum data can also be used to estimate  masses.   A  high-mass cluster progenitor will manifest as a dense ($n > 10^4$ cm$^{-3}$), cold ($T<\sim15$ K)  molecular cloud with sufficiently high mass ($>\sim 10,000$ \Msun) to form a high-mass star cluster. For a YMC progenitor, \citet{Bressert12} suggest that a mass of $>3\times10^4$ \Msun\, is required.  
Based on these criteria, \citet{Contreras17} identified one cold YMC precursor from the roughly 3,500 ATLASGAL clumps observed in the MALT90 survey.  In fact, this high-mass cluster precursor, AGAL331.029-00.431, actually consists of two ATLASGAL clumps: AGAL331.029-00.431 and AGAL331.034-00.419.  Its combination of cold temperature ($T_{dust} = 14$ K) and large mass ($M=41,000$ \Msun\, for the more likely far kinematic distance of 10 kpc;  $M=6,600$ \Msun\, for the less likely near kinematic distance of 4 kpc) are unique for dense clouds outside of the Galaxy's Central Molecular Zone.

In this paper we identify an additional candidate for a high-mass cluster precursor, \Pebble, or ``the Pebble''  (similar to the ``Brick,'' but with a smaller mass).  We show that  \Pebble\, is composed of four ATLASGAL clumps whose molecular line properties demonstrate that they all belong to single, coherent cloud.  In addition, despite its low dust temperatures, \Pebble\, has surprisingly large linewidths.  Unfortunately, the disambiguation of the near/far kinematic distance ambiguity remains difficult for \Pebble, resulting in an estimated mass of either 5,000 \Msun\, for the near kinematic distance or 27,000 \Msun\, for the far distance.   The larger mass would just qualify \Pebble\, as a potential precursor for a YMC.  If \Pebble\, will form a cluster with a star formation efficiency of $\sim 30$\%,  it would still form a cluster of mass 1,500 \Msun\, even if the smaller mass associated with the near kinematic distance is correct.  At either distance, \Pebble\, is an excellent candidate for a second cold high-mass cluster precursor outside of the CMZ.  

\section{Observations and Results}
To determine properties based on  the dust continuum emission from \Pebble, we follow the procedures of \citet{Guzman15} to analyze the 870 \um\, data from the ATLASGAL survey \citep{Schuller09} along with the 70 to 500 \um\, data from the HiGAL survey 
\citep{Molinari10}.  Using these procedures, \citet{Guzman15}  produced maps of the dust temperatures and gas column 
densities from the dust conitnuum spectral energy distributions.  
To analyze molecular line properties, we use the $J = 1-0$ rotational transition data from \hcopnt, \nthpnt, HCN, and HNC in the MALT90 Survey \citep{Jackson13, Rathborne16}.

\begin{figure}[!t]
\includegraphics{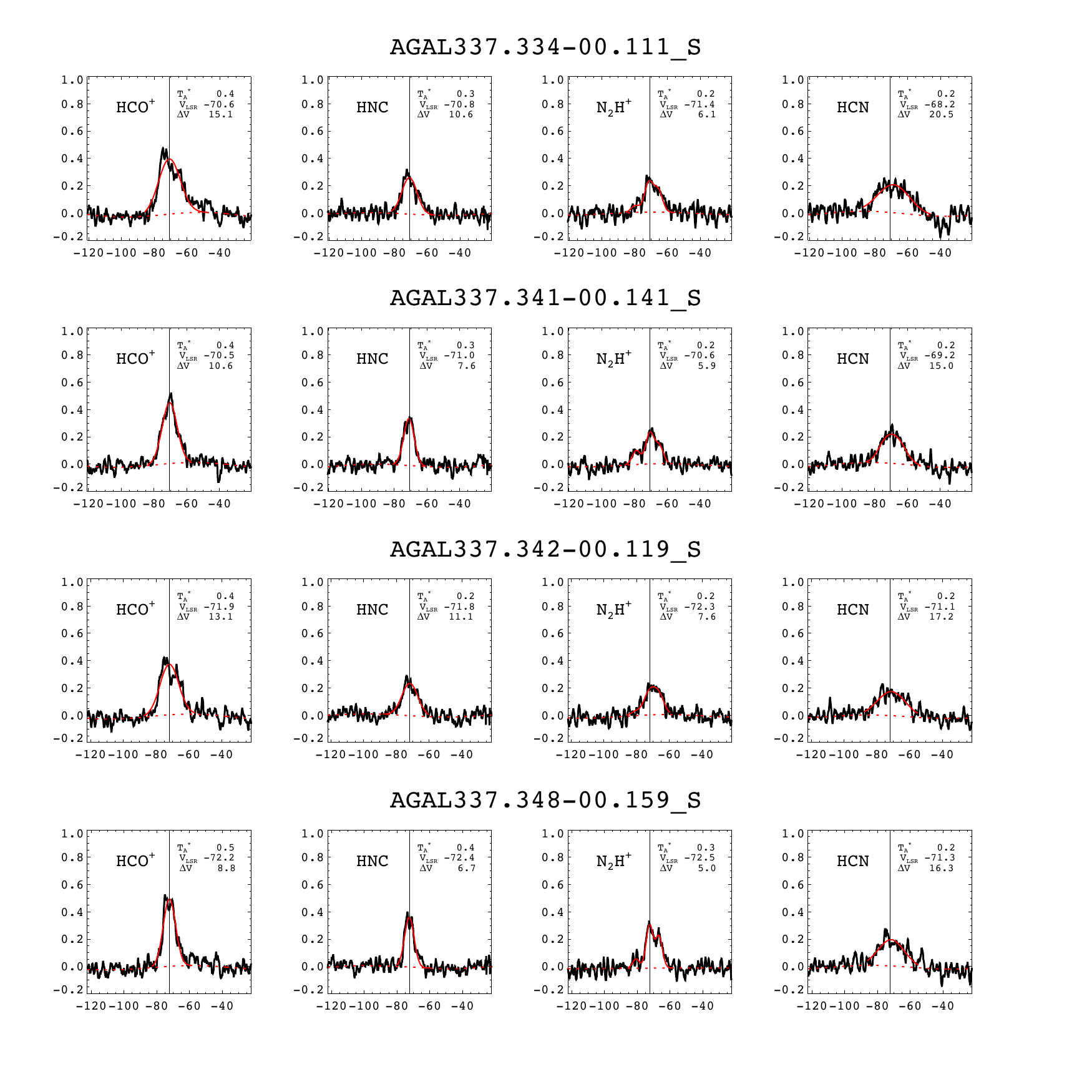}
\caption{Molecular line spectra of the 1-0 transitions of HCO$^+$, HNC, N$_2$H$^+$, and HCN toward the central positions of the four ATLASGAL clumps associated with \Pebble.  Gaussian fits to the spectra are shown superposed in the solid red lines, while the baseline is indicated by the dashed red line. The vertical lines mark the flux-weighted ``consensus velocity'' for each clump (see \citealt{Rathborne16}.)}
\label{Figure1}
\end{figure}

\begin{figure}[!t]
\includegraphics{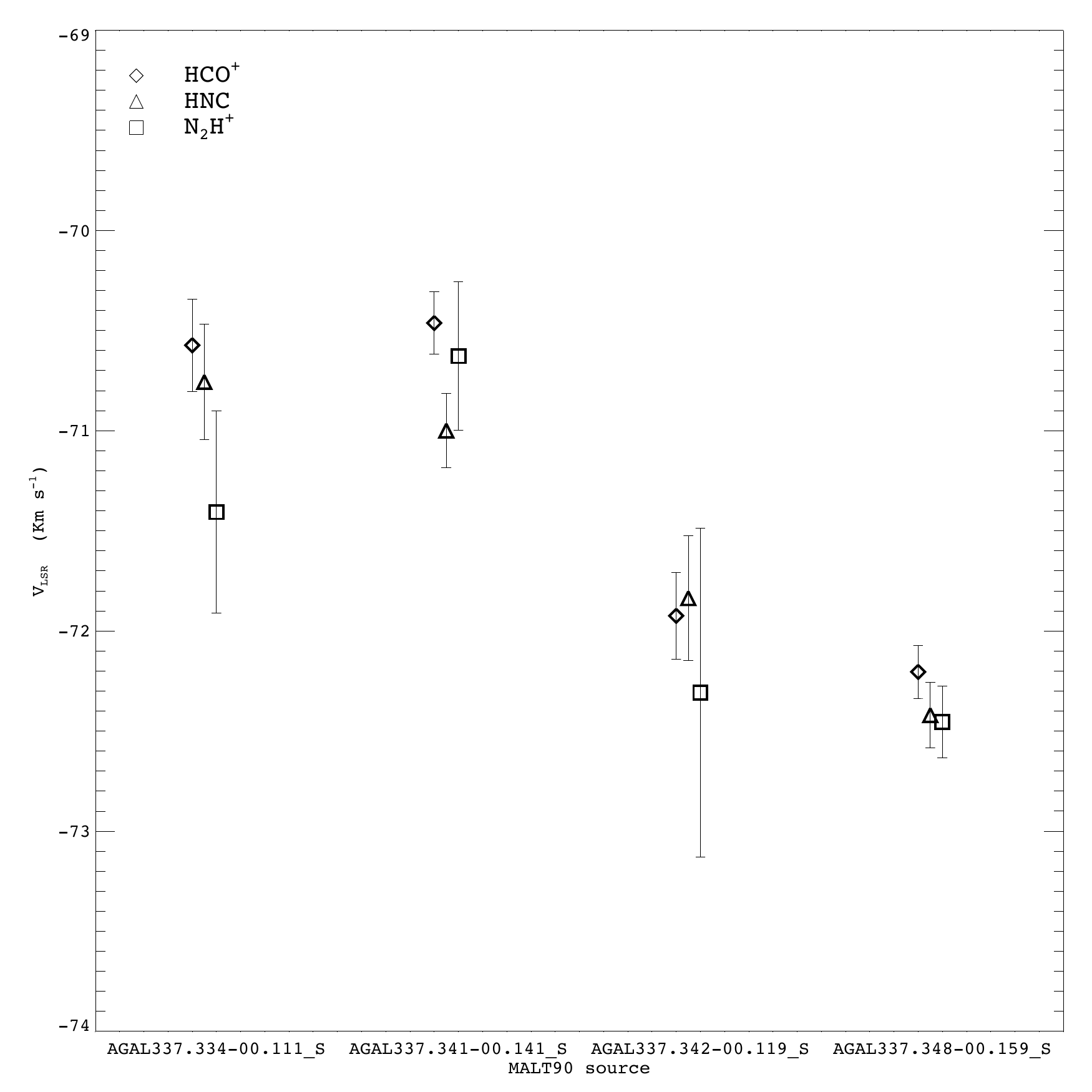}
\caption{The central velocity from Gaussian fits of the 1-0 transitions of HCO$^+$, HNC, and N$_2$H$^+$ toward the central positions of the four ATLASGAL clumps associated with \Pebble. }
\label{Figure2}
\end{figure}

\begin{figure}[!t]
\includegraphics{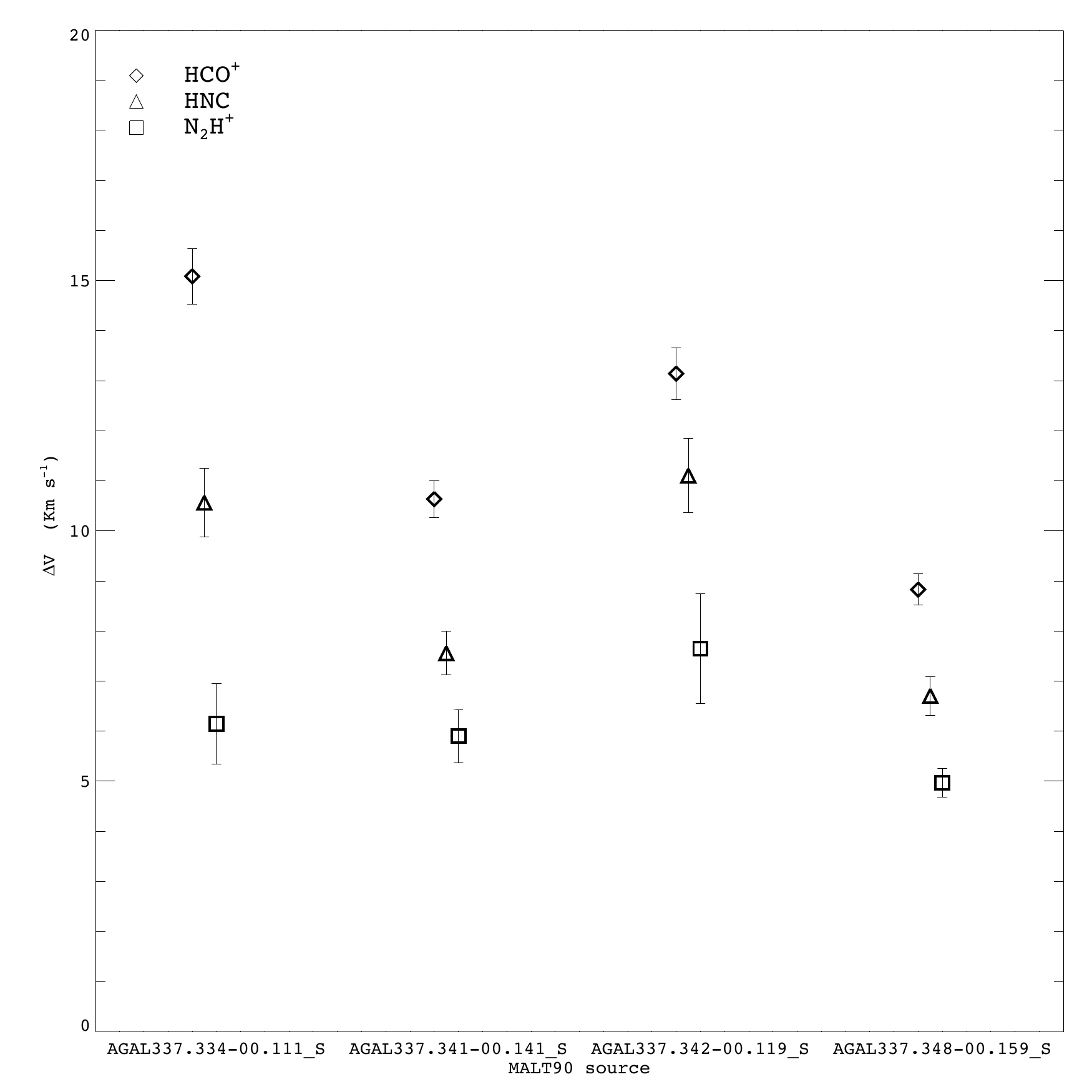}
\caption{The full-width half-maximum (FWHM)  linewidths from Gaussian fits of the 1-0 transitions of HCO$^+$, HNC, and N$_2$H$^+$ toward the central positions of the four ATLASGAL clumps associated with \Pebble. }
\label{Figure3}
\end{figure}

In the course of fitting Gaussians to the MALT90 spectra \citep{Rathborne16}, we noticed that four contiguous ATLASGAL clumps--- 
AGAL337.334$-$0.111, AGAL337.341$-$0.141, AGAL337.342$-$0.119, and AGAL337.348$-$0.159--- all have similar molecular line velocities 
and linewidths in \hcopnt, HNC, \nthpnt, and HCN $1-0$ emission (see Figures \ref{Figure1}, \ref{Figure2}, and \ref{Figure3}, and Table 1). 
Particularly striking are the unusually  large linewidths $\Delta V\sim5$ to 15 \kms\, (Fig. \ref{Figure3}, Table 1), much larger than the typical MALT90 linewidths of $\sim2$ to 3 \kms\, \citep{Rathborne16}.    In this paper ``linewidth" refers to the FWHM linewidth $\Delta V$, not the velocity dispersion $\sigma$. Note that because of the large optical depths and extreme blending of the HCN $1-0$ hyperfine components, only a single Gaussian component was fit to the HCN emission.  This will lead to overestimates of both the HCN LSR velocity and the linewidth. Because the HCN spectra are difficult to interpret, we exclude them from analysis here.   For the N$_2$H$^+$ $1-0$ line, however, the three main hyperfine components were well separated and therefore could be fit simultaneously.  Due to the smaller optical depths and less extreme blending of the hyperfine lines, we retain the N$_2$H$^+$ data for analysis.)

\begin{figure}[!t]
\includegraphics[width=0.32\textwidth,trim=35mm 50mm 50mm 50mm,clip=true]{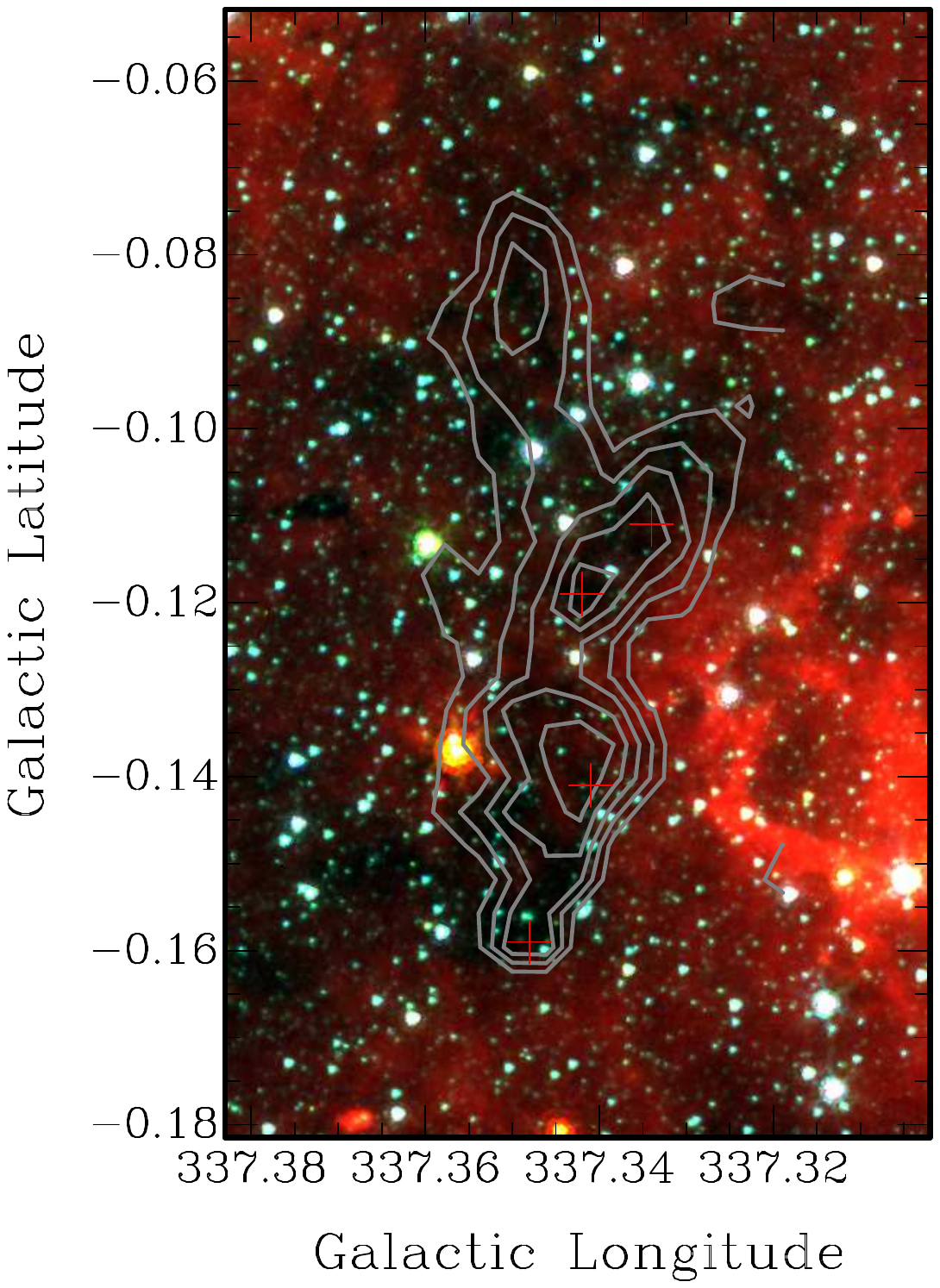}
\includegraphics[width=0.32\textwidth,trim=20mm 35mm 20mm 20mm,clip=true]{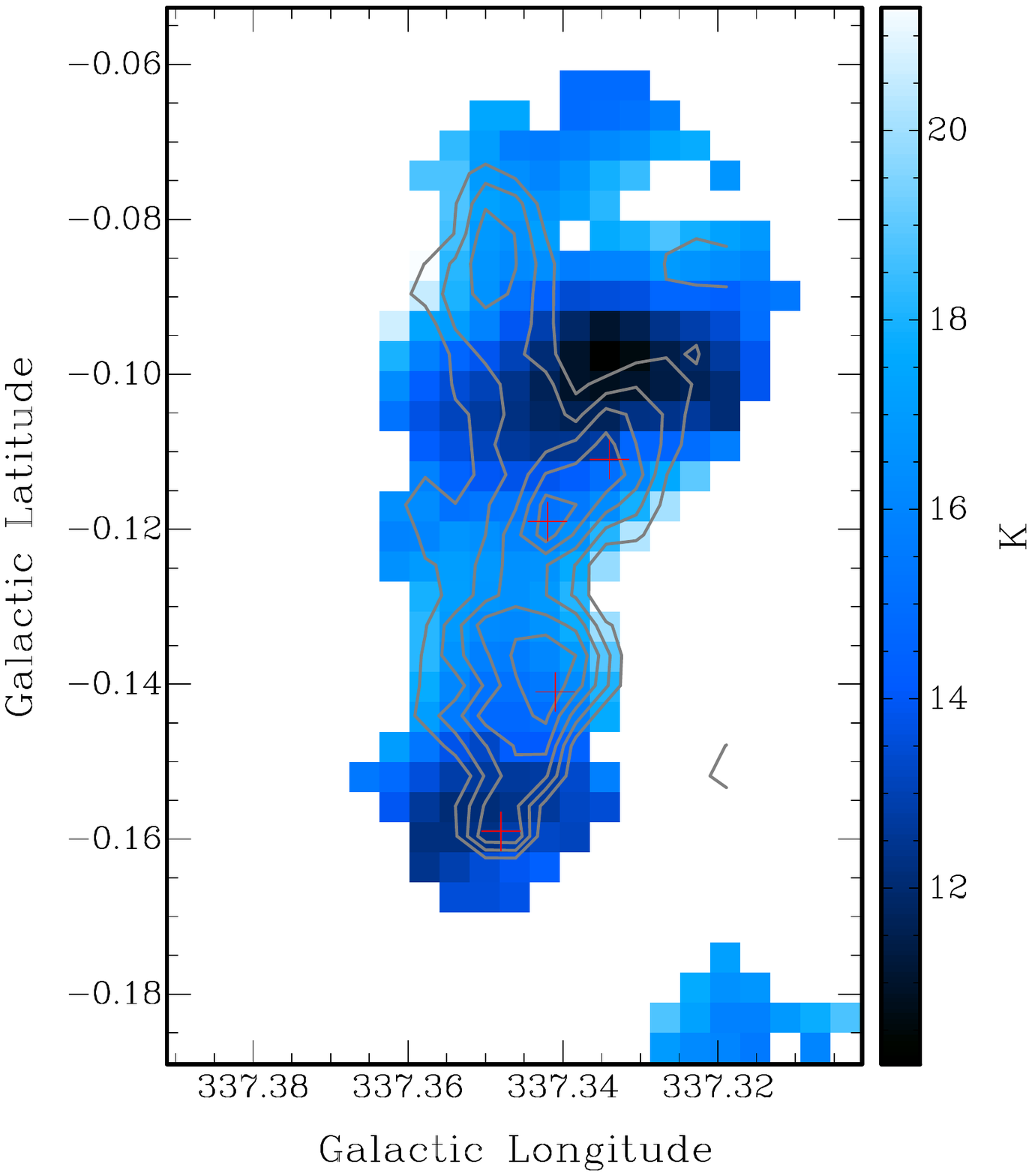}
\includegraphics[width=0.32\textwidth,trim=20mm 30mm 15mm 20mm,clip=true]{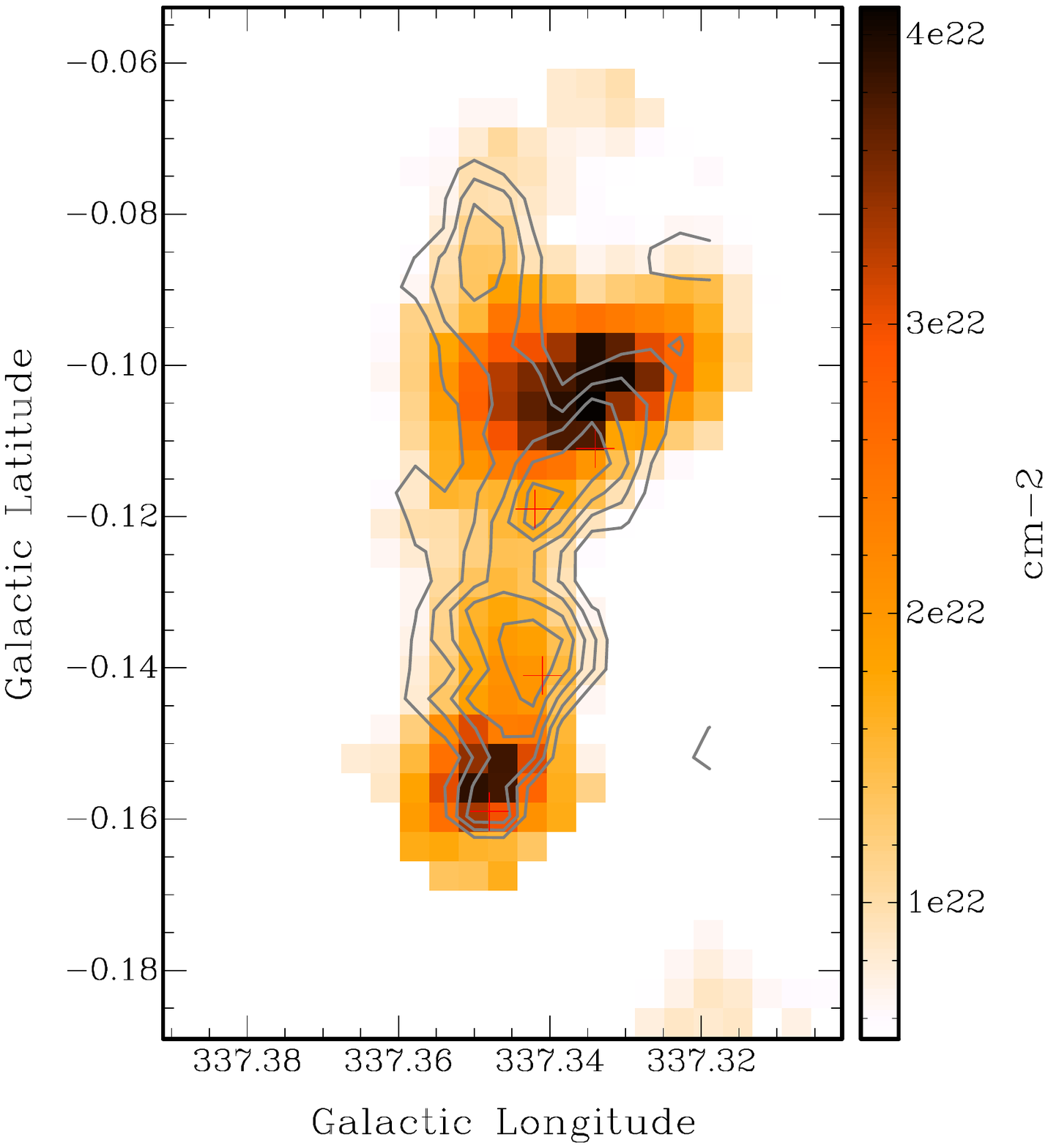}
\caption{Images of G337.342-0.119: (left) Spitzer/IRAC three color image with 3.6 \um\, emission in bllue,
4.5 \um\, emission in green, and 8.0\um\, emission in blue, (middle) the Herschel-derived dust
temperature, and (right) H$_2$ column density maps.  On all images, the contours are ATLASGAL 870\um\, dust
continuum emission, 50 to 90\% of the peak in steps of 10\%. }
\label{Figure4}
\end{figure}

Following the procedure of \citet{Guzman15}, we have  produced images of the dust temperature and the gas column density for the region containing all four clumps (Figure \ref{Figure4}). 
In this analysis of HIGAL \citep{Molinari10} and ATLASGAL \citep{Schuller09} submillimeter and far infrared data,  all images are convolved to a common angular resolution of 29.3$''$ and emission with spatial scales larger than 2.5$'$ is filtered out of the images.  Moreover, the analysis adopts the absorption coefficients $\kappa$ from models of silicate-graphite dust grains with a coagulation age of $3\times10^4$ yr from \cite{Ormel11} and a gas-to-dust mass ratio of 100.  The absorption coefficient $\kappa$ is a funciton of function of frequency (see Figure 6 of \citealt{Ormel11}). At a wavelength of 100 $\mu$m, $\kappa \sim 100$ cm$^2$ g$^{-1}$.  Dust temperatures throughout the region are low ($T_{dust}$ =12 to 16 K), and smaller than those of typical MALT90 target clumps, even for ``quiescent'' clumps  with no obvious star formation activity, which average $T_{dust} = 16.8 \pm 0.2$ K \citep{Guzman15}.  More evolved clumps show even higher dust  temperatures: $18.6 \pm 0.2$ K for ``protostellar,'' $23.7 \pm 0.2$ K for ``\hii\, region,'' and $28.1 \pm 0.3$ K for ``photodissociation region'' clumps \citep{Guzman15}.

We have estimated kinematic distances for each of the four clumps (see \citealt{Whitaker17} for details).  Based on the observed MALT90 molecular line velocities, the near kineamtic distance toward these clumps is 4.7 kpc, and the far kinematic distance is 11.0 kpc.   In principle, the kinematic distance ambiguity can be resolved by 21 cm H I measurements, but the results are ambiguous.  For AGAL337.342$-$0.119 and AGAL337.341$-$0.141, the algorithm of \citet{Whitaker17} slightly favors the near kinematic distance; however, for the two remaining clumps, AGAL337.342$-$0.119 and AGAL337.348$-$0.159, the algorithm slightly favors the far kinematic distance.  In no case is the probability of a correct near/far assignment greater than 0.8.  The distance estimation technique of \citet{Foster12} that employs infrared data on star counts and colors slightly favors the near kinematic distance.   If the the near kinematic distance is correct, however, one might expect to see significant extinction in the mid-IR such as that found in Infrared Dark Clouds (IRDCs) with similarly high column densities and cold temperatures.  Such mid-IR extinction, however, is entirely absent (Fig. 4).  Thus, the kinemitic distance ambiguity for \Pebble\, remains unresolved.  Consequently, when physical parameters depend on distance, values are presented for both the near and far kinematic distances.

\begin{figure}[!t]
\includegraphics[width=1.0\textwidth]{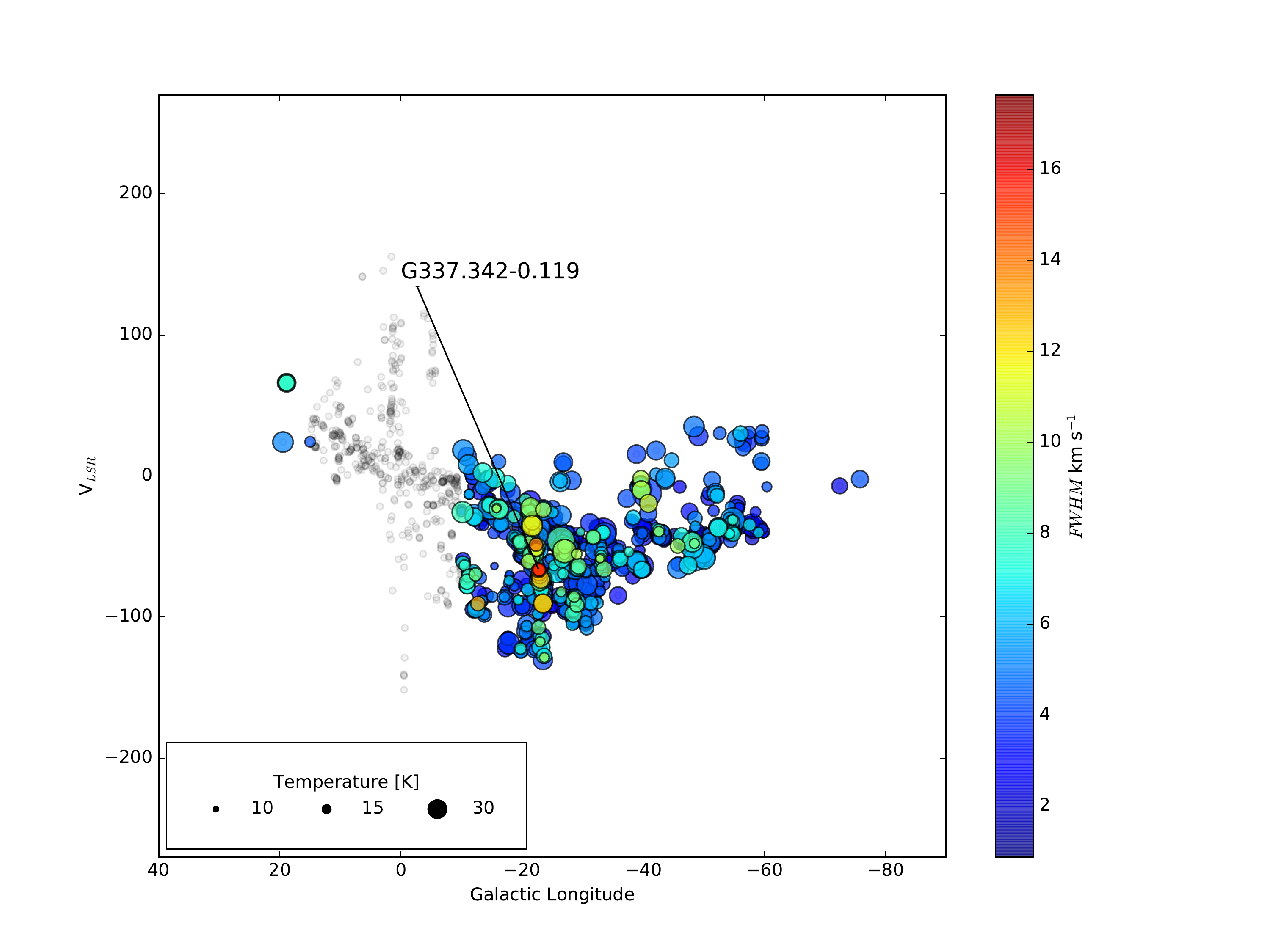}
\caption{A longitude-velocity ($l-V$) diagram showing the location of MALT90 target clumps, coded by FWHM linewidth ($\Delta V$) of the \hcop\, line (color) and dust temperature (symbol size).  All MALT90 sources meeting the following criteria were included for further analysis: the sources must lie outside of the Central Molecular Zone (sources with $350^\circ < l < 360^\circ$ and $0^\circ < l < 15^\circ$, indicated by the gray points, are excluded); each has a significant ($>4\sigma$) detection of \hcop, a significant dust temperature determination (Guzm{\'a}n et al. 2015), and a line shape best modeled by a single Gaussian with no high residuals after subtracting a Gaussian fit (see Rathborne et al. 2016 for details).  Of the 1,060 MALT90 sources meeting these criteria, \Pebble\, has a unique combination of cold dust temperature and large linewidth.}
\label{Figure5}
\end{figure}

Among clumps in the Galactic disk, the four clumps comprising \Pebble\, have a unique combination of cold dust temperatures and large linewidths.   Figure \ref{Figure5}  displays a longitude-velocity diagram for MALT90 sources detected in \hcopnt. To ensure that we are examining clumps in the Galactic Disk and excluding the Central Molecular Zone  sources with large turbulent linewidths, we have excluded sources with $350^\circ < l < 360^\circ$ and $0^\circ < l < 15^\circ$.   In addition, to be included in Figure 5, each MALT90 source must have a significant ($>4\sigma$) detection of \hcop, a significant dust temperature determination \citep{Guzman15}, and a line shape best modeled by a single Gaussian with no high residuals after subtracting a Gaussian fit (see \citealt{Rathborne16} for details).  Of the original 3,250 MALT90 sources, 1,060 sources meet all of these criteria. In Figure \ref{Figure5}, the size of the circles represents the dust temperature, with smaller circles having smaller dust termperatures, and the color of the circles represents the linewidths.  \Pebble\, clearly stands out for its unique combination of cold dust temperature and large linewidth.

\section{Discussion}

\subsection{The global properties of a single cloud}
The four ATLASGAL clumps comprising \Pebble\, present a unique combination of cold dust temperatures and large molecular linewidths.  One possibile explanation for these large linewidths is that two clouds with more typical, smaller linewidths are superposed along the line of sight, and the line emission from the two clouds blends together to mimic a single line of larger linewidth.
We consider this explanation unlikely.  If two clouds were superposed along the line of sight, their extents and positions would not perfectly align.  Thus, one would expect spatial gradients
in the LSR velocities, in the linewidths, and in the dust temperatures at positions where the emission from one cloud or the other dominates.  Yet, for \Pebble, the LSR velocities and the linewidths are remarkably similar across its entire extent, with no sudden discontinuities (see Fig. 6). (The smooth velocity gradient in the direction of Galactic longitude evident in the HNC and HCO$^+$ images cannot account for the observed linewidths. The linewidth due to beam-smearing of a source with a gradient can be approximated as $\Delta V_{smear} \sim dV/d\theta \times \theta_{beam} \sim 1$ \kms\, for the observed gradients and the 38$''$ Mopra beam.  This is much smaller than the observed linewidths.)  In order for two clouds to mimic the properties of the \Pebble, both clouds would need to be perfectly aligned and have exactly the same angular extent. We concude that the \Pebble\, is a single, coherent cloud, albeit with significant substructure.

If we combine the four ATLASGAL clumps into a single object, we can determine its overall properties.  Since many of these properties depend on distance but the near/far  kinemtic
distance ambiguity cannot be definitively resolved, we calculate two sets of properties, one set corresponding to the near kinematic distance of 4.7 kpc, and a second set corresponding to the far kinematic distance of 11.0 kpc.  These properties are presented in Table 2.  We determine the mass by integrating the column density maps derived from the dust continuum (Fig. 4) over the area: $M = \int N \mu m_H dA$.  Here $N$ is the column density, $\mu$ is the molecular weight (assumed to be 2.6),  $m_H$ the mass of a hydrogen atom, and $A$ the area.  
The radius $R$ is found by the angular extent $\theta$: $R =\frac{1}{2}  \theta D$, where $D$ is the kinematic distance.  Here we estimate the angular extent $\theta$ to be roughly the size of the ~50\% ATLASGAL flux contour: $\theta_{maj} \sim 0.08^\circ$ and  $\theta_{min} \sim 0.02^\circ$.  
The average density is found by $n = (3M)/(4\pi \mu m_H R_{maj} R_{min} R_{los})$ where $R_{maj}$ is the radius of the major axis, $R_{min}$ is the radius of the minor axis, and $R_{los} = (R_{maj} + R_{min})/2$ is the average projected radius.  
Here we assume that the cloud can be approximated by a triaxial elliipsoid and that the radius in the line-of-sight direction, $R_{los}$ is the average of the radius of the major and minor axes.  If the cloud is in fact filamentary and $R_{los}$ is better approximated by $R_{min}$, the actual densities will be a factor of 2.5 higher than the reported densities.
We determine the virial mass by $M_{vir}=(5\sigma^2R_{av})/G$, an approximation valid for a uniform density distribution.  Here $\sigma$ is the velocity dispersion, and we assume a value of $\Delta V = 10$ \kms. The virial parameter is found from $\alpha = M_{vir}/M$.  Finally, the dust temperatures are derived from the procedures described in Guzm{\'a}n et al. (2015).  

Whether it lies at the near or the far kinematic distance, \Pebble\, has the large mass and cold dust temperature consistent with a cold progenitor of a  high-mass cluster (and possibly a YMC).  If it lies at the far distance, \Pebble\, has very nearly the minimum
mass predicted for it to be a YMC progenitor ($\sim 30,000$ \Msun).  If instead it lies at the near distance, its mass is 5,000 \Msun, sufficient to produce a star cluster with stellar mass $M_* \sim 1,500$ \Msun\, (if the star formation efficiency is 30\%).
Its size $R_{av} = 2$ to 5 pc is on the higher end of the size distribution of MALT90 clumps \citep{Contreras17} and is comparable to that of the Brick (3 pc). This large size may suggest that only extremely large precursor clumps contain sufficient mass to form a high-mass cluster or a YMC.  Possibly collapse motions in \Pebble\, will lead to a smaller cloud as it evolves.  In the simplest approximation of a cloud of fixed mass, for a {\bf plausible} infall speed of $\sim 0.2$ \kms, a cloud can reduce its radius by 1 pc in $\sim 5 \times 10^6$ yr.  

The average densities for \Pebble\ are estimated to be 1,400 and 3,300 \pcc\, for the far and near kinematic distances, respectively, under the assumption that the line of sight radius $R_{los}$ is the average of $R_{maj}$ and $R_{min}$, and 3,500 and 8,300 \pcc\, for the far and near kinematic distances, respectively, under the assumption of a more filamentary geometry where $R_{los} = R_{min}$.  These densities are an order of magnitude higher than that of typical giant molecular clouds, for which typical denisties are $\sim$300 \pcc\, (e.g., \citealt{Solomon79}).  Emission from lines with high critical densities such as the HCO$^+$, N$_2$H$^+$, and HNC $1-0$ lines detected toward \Pebble\, is often interpreted as requiring densities $n>n_{crit} \sim 10^5$ \pcc, but radiative trapping can reduce the required densities for detectable emission to a few $10^3$ \pcc\, \citep{Shirley15}.  If \Pebble\, has clumpy structure on subparsec scales, the actual densities will be larger than the average density.  We conclude that \Pebble\, has a sufficiently large density to be a candidate high-mass cluster progenitor.

\subsection{Virial Equilibrium}
The virial parameter, $\alpha$, for \Pebble\,  is large, $\alpha \sim 4$ to 9. These values for $\alpha$ for the entire Pebble also match the average of the values of $\alpha$ for each of the four ATLASGAL clumps comprising the Pebble.  For gravitational collapse to ensue, theory suggests that $\alpha < 1$, although the simulations of \cite{Ballesteros2017} show that the use of the uniform, spherical approximation would lead to estimates closer to $\alpha \sim 2$ for collapsing clouds.
Since a cold molecular precursor to a high-mass cluster must eventually collapse in order to form a bound cluster, it is puzzling that, at first glance, \Pebble\, appears to be gravitationally unbound.  If \Pebble\, is to collapse, or is to remain in an equilibrium state, either additional forces besides gravity and turbulent pressure are important and act to prevent the cloud from expanding, or our estimate of $\alpha$ is incorrect.

We first consider additional forces at play besides those due to gravity and tubulent pressure.  For example, magnetic forces are often speculated to be important during the high-mass star formation process, but their exact role remains uncertain. Although magnetic fields can possibly provide a significant outward pressure that can support a cloud against collapse, it is difficult to imagine  how magnetic fields with any plausible geometry could provide a significant {\it  inward} force to  enhance cloud collapse or confinement.    

In addition to magnetic fields, external pressure is often thought to play a role in cloud confinement. In order for a uniform cloud in equiliibrium, for which turbulent pressure and gravity are the dominant internal forces, to be confined by an external pressure $P_{ext}$, it can be shown by applying the generalized virial theorem that the following relation holds:
$$P_{ext} = {{3\sigma^2 M}\over{4\pi R^3}} ({1 - {{1}\over{2\alpha}}})~~~~.$$   For the values of $\alpha$ estimated for \Pebble, external pressures of $P/k \sim 6\times 10^5$ and $1 \times 10^6$ cm$^{-3}$ K are required to confine the cloud in equilibrium for the far and near kinematic distances, respectively.  If this pressure is in the form of thermal gas pressure, these values seem implausibly large for either more diffuse molecular gas with $n \sim 10^3$ \pcc\, or atomic gas with $n\sim 1$ \pcc, because such values would require implausibly large temperatures: $T \sim 1,000$ K for the molecular gas or $T \sim 10^6$ K for the atomic gas.    Alternatively, ram pressure from a shock with a relative velocity $v_s \sim \Delta V$ could conceivably provide enough pressure to confine the cloud, but unless the shock is maintained with an approximately constant flow velocity over time, any cloud confinement via ram pressure would likely be transient. A cloud-cloud collision, the expansion of gas from a star-forming region, or supernova explosions are possible causes of transiently enhanced linewidths.  If an impulsive event only recently enhanced the turbulent linewidths in \Pebble, it should dissipate on a timescale $t \sim R/\Delta V \sim 0.75$ Myr.
If protostellar cores have already formed and begun to collapse they may form high-mass stars that live and die on timescales comparable to the cloud expansion timescale.   Thus, the dissipation of the cloud could conceivably have little effect on the resultant cluster if star-formation has already been triggered and commenced through its earliest phases.

It is of course possible that \Pebble\,itself  is a transient cloud, but exactly how \Pebble\, has accumulated so much mass despite its large turbulent velocity dispersion is also difficult to understand.  Since the dust temperatures are low and the infrared images show little evidence for embedded star formation activity, embedded star formation is unlikely to contribute the energy that generates the observed large linewidths. 
Possibly star formation may have already begun but, if so, the stars are too faint or young to have significantly heated large portions of the clump. Alternatively, the large linewidths may result from the influence of nearby star formation, by which an initially quiescent, narrow-linewidth cloud has had turbulent energy injected into it from a neighboring active region.  This possibility is discussed below.

\subsection{Possible Errors in the Estimate of the Virial Parameter $\alpha$}

An alternative explanation for apparent values of the virial parameter $\alpha > 1$, which suggest expansion rather than collapse, is an erroneous estimate for $\alpha$.  For example, if the linewidths for \Pebble\, are overestimated, then $\alpha$ would also be overestimated.  One possibility for such an overestimate of the linewidths is that large optical depths lead to saturated, broadened lines whose measured width exceeds the true velocity dispersion.  In fact, the linewidths derived from \hcop\, and \nthp\ differ  by roughly a factor of two, with the optically thicker \hcop\, line having the larger linewidths.  Yet, if line broadening due to large optical depths in \hcop\, were responsible for the large observed linewidths, the optical depths for \hcop\, would be so large that we would expect to easily detect the \htcop\, line.  However, \htcop\, was not detected in MALT90.  The larger linewidths for the optically thicker lines may instead arise from their ability to trace regions of lower column density but larger turbulent linewidths, perhaps related to the requirements for their chemical production. Other possibilities for errors in deriving $\alpha$ are incorrect estimates of the cloud radius, incorrect gas masses derived from the dust continuum due to errors in assumptions about  dust temperatures, dust opacities, or gas to dust ratios, or incorrect assumptions about cloud structure (e.g., centrally peaked density distributions rather than a uniform distribution), geometry (cylindrical collapse versus spherical collapse) or the absence of larger scale bulk motions.  With realistic assumptions, such errors probably lead to errors of factors of at most a few. (See \cite{Guzman15} for a detailed  discussion of the errors related to the derivation of parameters from the dust continuum, namely dust temperature and column density.) Thus it seems implausible that our estimates of $\alpha$ are consistent with true values of $\alpha<1$.  

\subsection{Galactic Environment}

The unique properties of \Pebble\, suggest that it may have a special location in the Milky Way.  Its broad linewidths are reminiscent of the active star formation region W43 in the Galaxy's first quadrant.
A number of studies have suggested that W43's extreme star formation activity arises from its location at the near end of the Galactic Long  Bar \citep{Luong11, Eden12, Zhang14}. It is interesting to speculate that the extreme properties of \Pebble\, may also arise from a similar location at the far end of the Long Bar.   If the Long Bar has a length
$L$ and an  inclination angle $i$ with respect to the line of sight to the Galactic Center, then the longitude of the end of the Long Bar at positive Galactic longitude, $l_+$ will in general differ 
from the longitude of the end of the Long Bar at negative Galactic longitude, $l_-$, due to projection.  The two longitudes are related by 
$$\tan{l_\pm} = {\sin{i} \over{{{2R_0}\over{L}}\mp \cos{i}}}$$
Here $i$ is the inclincation angle, $R_0$ the distance between the Sun and the Galactic Center, and $L$ the length of the Long Bar.  Figure \ref{Figure7} shows the geometry.  The length of the Long Bar has been estimated to be $8.8$ kpc  (for a Galactic Center distance from the Sun $R_0 = 8.5$ kpc) and its inclination {$\sim43^\circ$} \citep{Hammersley00, Benjamin05, Lopez07}.  The adoption of these values places the near end of the Long Bar at a Galacitic longitude $l_+ = 29.8$ deg, approximately the location of W43, and the far end at $l_- = -14.7$ deg.  Since the actual longitude of \Pebble\, is $l = -22.7$ deg, the location of \Pebble\, at the far end of the Long Bar is impossible, unless the Long Bar parameters are badly in error.  We conclude that \Pebble\, is unrelated to the Galaxy's Long Bar.  Instead, \Pebble\, is located in the Galactic disk at either the near or far kinematic distance (see Figure \ref{Figure7}).  The properties of \Pebble, therefore, do not arise from a unique Galactic location.

Although \Pebble\, is not associated with the end of the Long Bar, its Galactic environment may in fact be atypical.  Figures \ref{Figure8} and \ref{Figure9} show that \Pebble\, is located within a larger ridge of molecular gas that contains a large, active star-forming complex.  The gas in this ridge, traced by ATLASGAL and GLIMPSE continuum and MALT90 molecular line emission, have similar LSR velocities and linewidths as those found in \Pebble.  Thus, the entire molecular complex is coherent and exhibits large linewidths throughout. Moreover, a clear velocity gradient mostly in the direction of Galactic longitude extends smoothly from \Pebble\, through the star-forming complex.  This association suggests that \Pebble, may have formed from, or have been influenced by, the action of star formation in this large, active, star-forming complex.  

The exact reason for the enhanced linewdths in \Pebble\, and its associated star-forming complex remain unclear.  One possibility is that they arise from large-scale, global collapse.  For free--fall collapse where only gravity dominates,  infall motions of roughly 2 to 3 \kms\, might be expected for the mass and radius of \Pebble\, and the associated star-forming region.  However, given the large turbulent pressure, the actual collapse speeds should be much smaller.  Alternatively, energy injection from the combined gas flows (expanding H II regions, outflows, and/or supernova explosions) produced by the star-forming complex may have produced larger than usual linewidths throughout the region.  This star forming complex may also be physically expanding into \Pebble, which might account for the observed velocity gradient along the direction connecting \Pebble\, with the complex. 

\subsection{Is \Pebble\, a protocluster?}
\Pebble\, has the large mass and cold temperature expected for a young precursor to a high-mass star cluster, yet the question remains whether the large turbulent linewidths will prevent it from actually forming a high-mass cluster. Indeed, \cite{KrumholzMcKee05} suggest that a high degree of turbulence supresses star formation, and the large virial parameter of \Pebble\, would seem to preclude gravitational collapse.  Nevertheless, high-mass clusters such as the Arches and Sgr B2 have clearly formed in the turbulent Galactic Center region where molecular clumps have similarly large or even larger  linewidths.  Thus, a high degree of turbulence apparently does not always preclude high-mass cluster formation.  The physical conditions of clumps in their earliest stages of high-mass cluster formation remain unclear, and it is possible that these young, cold precursor clumps could be highly turbulent.  Although increased turbulence may act to suppress the overall star formation efficiency of a clump, it may also raise both the threshold density for star formation as well as the mass accretion rate once the protostar begins to accumulate mass.   At this point, the future evolution of \Pebble\, is speculative, but the possibility remains that it could be a cold precursor to a high-mass cluster.  Future high-resolution observations of its small-scale structure may determine whether pre-stellar cores have formed.

\section{Conclusions}
We have used the ATLASGAL, HiGAL, and MALT90 surveys to 
search for new examples of cold, high-mass cluster progenitors.  From over 3,000 fourth quadrant ATLASGAL clumps surveyed by MALT90, 
we have identified a new candidate: \Pebble, or ``the Pebble.''  \Pebble\, consists of four ATLASGAL clumps with very similar molecular line properties.
\Pebble\, has a unique combination of low dust temperatures ($\sim 12$ K) and large linewidths ($\Delta V \sim 6$ to 20 \kms).  The homogeneity of the 
molecular line parameters demonstrates that the four ATLASGAL clumps are in fact parts of a larger, coherent cloud.  The kinematic distance for \Pebble\,
is 4.7 kpc at the near kinematic distance, and 11 kpc at the far kinematic distance, corresponding to  a mass of 5,000 or 27,000 \Msun, respectively.
The near/far kinematic distance ambiguity has not been resolved, however.

The extraordinarily large linewidths for a source outside the Galaxy's Central Molecular Zone suggest that turbulence plays an important role in the cloud's
structure and energetics.  The virial parameter $\alpha$ is much larger than unity, and suggests that the cloud is either transient or confined by forces
in addition to gravity.  It is possible that such a large degree of turbulence is associated with cold, high-mass cluster progenitor clumps. The Galactic longitude of \Pebble\, 
is inconsistent with a location at the far end of the Galaxy's Long Bar.  Instead, \Pebble\, appears to be associated with a large star-forming complex which has many clumps with larger than usual linewidths.
\Pebble\, is one of the most massive, cold, dense molecular clouds known.  Further study of its detailed substructure may help to shed light on the initial conditions of 
the formation of high-mass clusters.

{\it Acknowledgements}:  JMJ gratefully acknowledges funding from the US National Science Foundation AST-0808001, the distinguished visitor program from the CSIRO, and from the University of Newcastle.  We thank the anonymous referee for a thorough reading of the original manuscript and for making several important suggestions that have greatly improved the paper. This research was conducted in part at the SOFIA Science Center, which is operated by the Universities Space Research Association under contract NNA17BF53C with the National Aeronautics and Space Administration.




\clearpage

\clearpage

\clearpage

\clearpage

\clearpage

\clearpage
\begin{figure}[!t]
\includegraphics[width=1.0\textwidth]{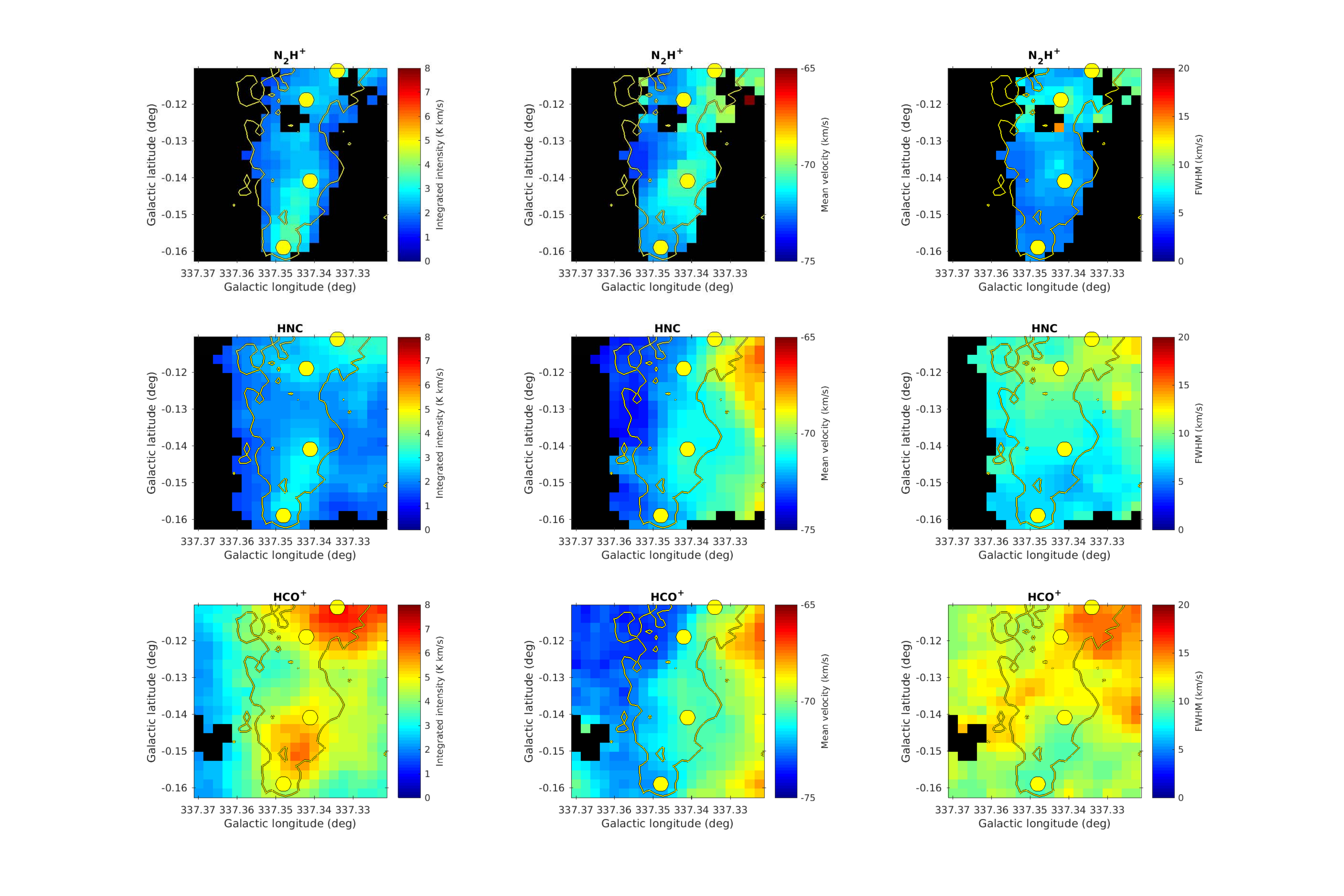}
\caption{Maps of the \nthp, \hnc, and \hcop\, emission from \Pebble, showing (left) the integrated intensity, (middle) the LSR velocity in \kms, and (right) the linewidth $\Delta V$ in \kms of the molecular line emission.  Pixels with a signal-to-noise ratio less than 10 in the integrated intensity are colored black. The x-axis is Galactic longitude in degrees and the y-axis Galactic latitude in degrees. The yellow  circles indicate the positions of the four ATLASGAL clumps comprising the ``Pebble.'' The contours represent the ATLASGAL emission at a flux level of 0.2 Jy.  The lack of large gradients or spatial discontinuities suggest that \Pebble\, is a single cloud and not the chance superposition of unrelated clouds.}
\label{Figure6}
\end{figure}

\clearpage
\begin{figure}[!t]
\includegraphics[width=0.8\textwidth]{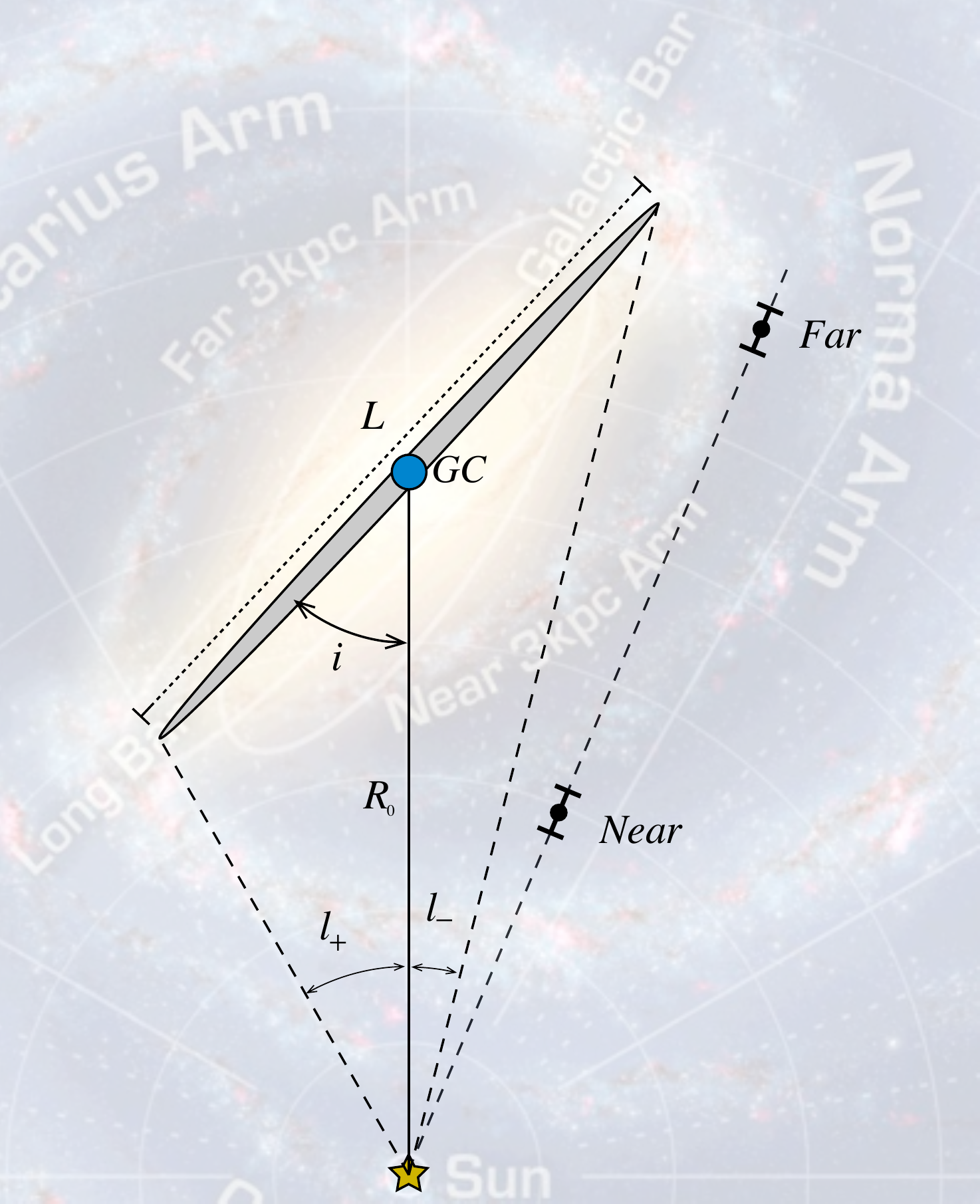}
\caption{The geometry of the Galaxy's Long Bar, superposed on a notional image of the Galaxy's structure \citep{Benjamin05}.  Here ``GC'' and ``Sun'' mark the locations of the Galactic Center and the Sun, respectively, $L$ is the length of the Long Bar, $R_0$ the distance to the Galactic Center from the Sun, $i$ the inclination angle with respect to the line connecting the Sun and the Galactic Center, $l_+$ the Galactic longitude of the near end of the Long Bar, and $l_-$ the Galactic longitude of the far end of the Long Bar.  "Near" and "Far" mark the near and far kinematic distances of \Pebble\, \citep{Whitaker17}.}
\label{Figure7}
\end{figure}

\clearpage
\begin{figure}[!t]
\includegraphics[width=1.0\textwidth]{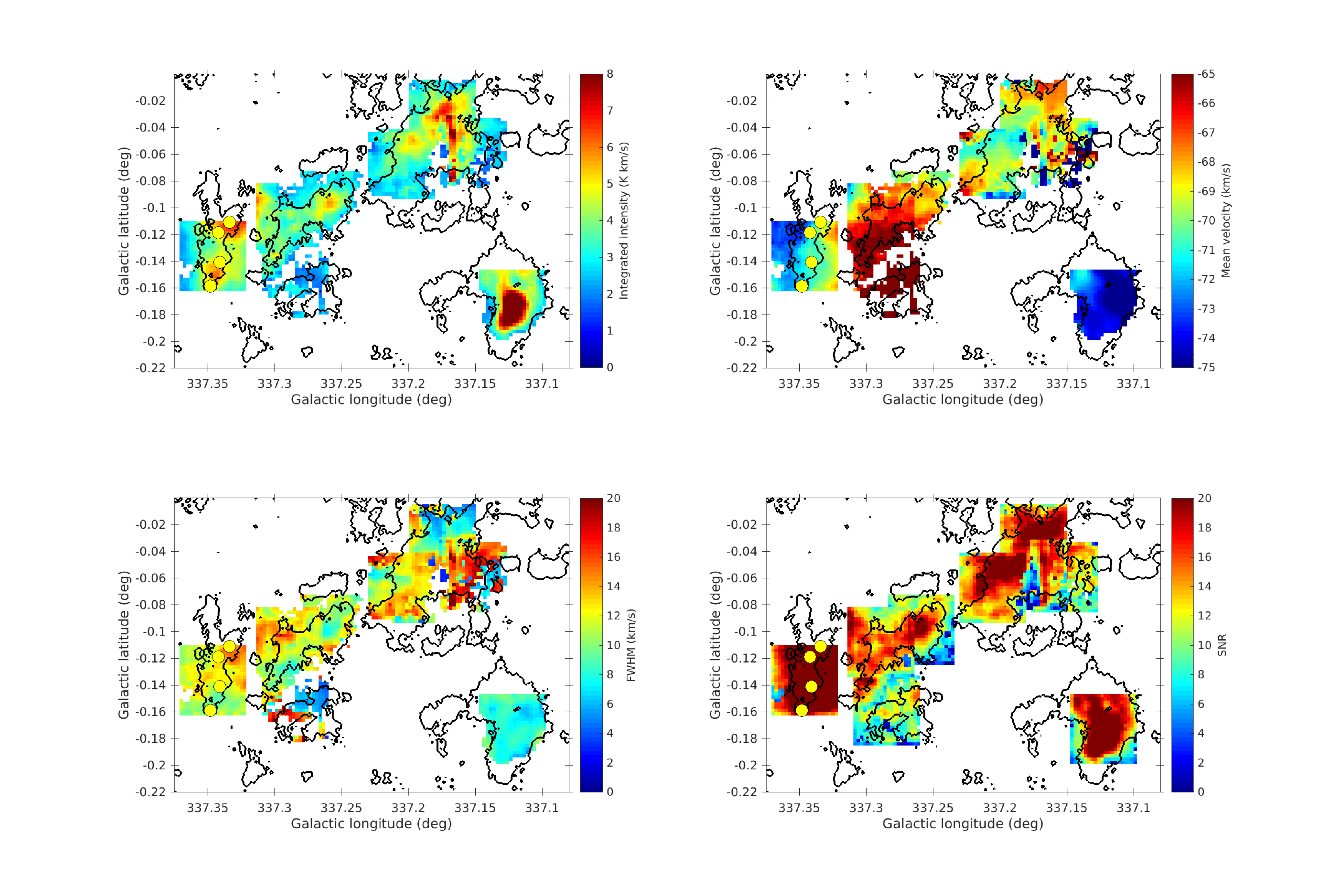}
\caption{Maps of MALT90 data for the \hcop\, line in mulitple fields including and nearby the position of \Pebble: (top left) Integrated intensity, (top right) LSR velocity in \kms, (bottom left) FWHM linewidth $\Delta V$ in \kms, and (bottom right) the signal-to-noise ratio for the integrated intensity.  The yellow circles indicate the positions of the four ATLASGAL clumps comprising the ``Pebble."   The contours indicate ATLASGAL 870 \um\, emission at flux level 0.2 Jy.  \Pebble\, is associated with a large ridge of molecular gas with larger than typical linewidths.}
\label{Figure8}
\end{figure}

\clearpage
\begin{figure}[!t]
\includegraphics[width=1.0\textwidth]{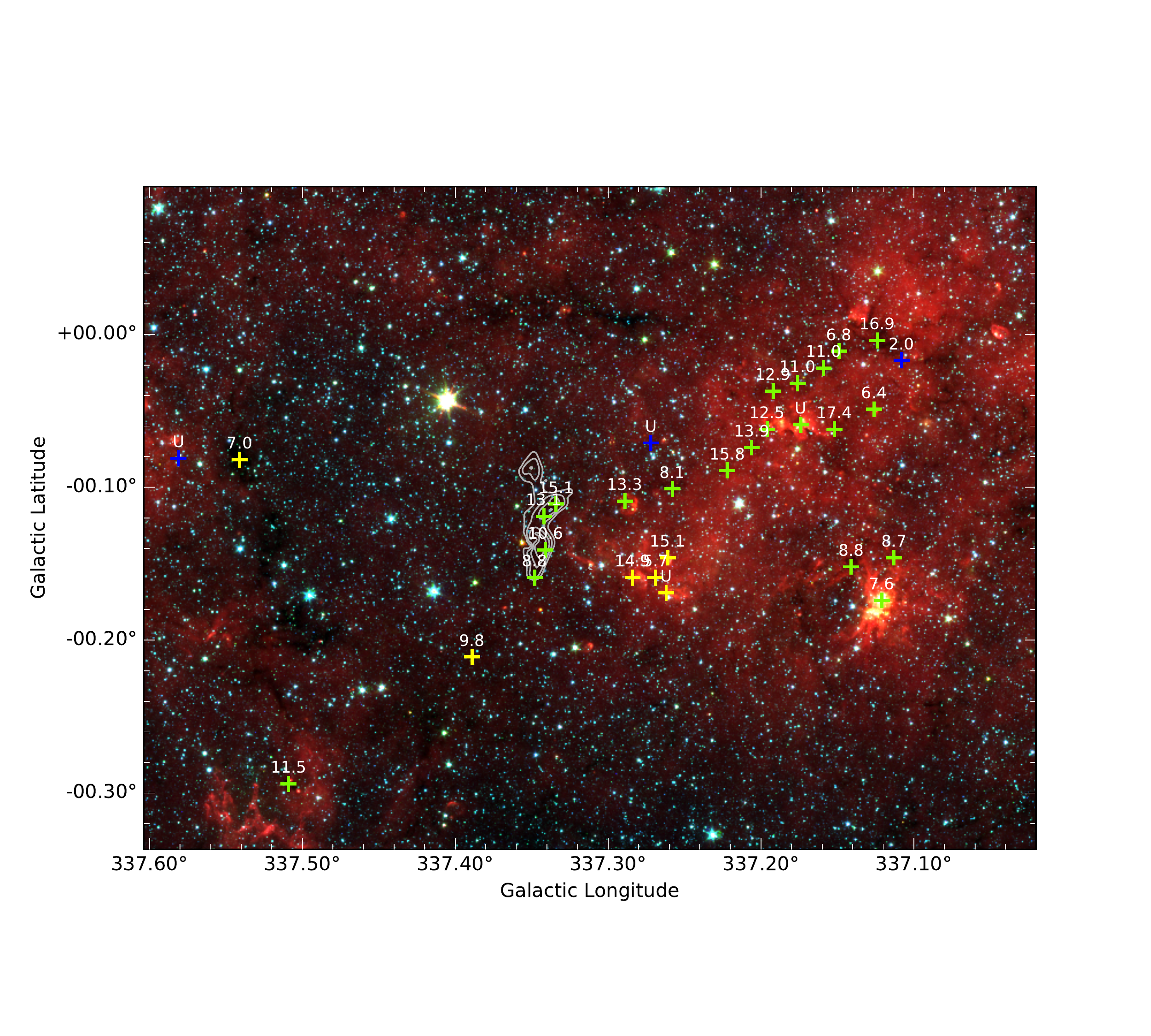}
\caption{A GLIMPSE image of the vicinity of the Pebble, with blue 3.6 \um, green 4.5 \um, and red 8.0 \um.  Contours indicate the ATLASGAL 870 \um\, image of dust associated with \Pebble. Superposed on the image are plus signs that indicate the positions of several ATLASGAL clumps in the vicinity of \Pebble, and the values of the FWHM linewidth of the \hcop\, line in \kms, derived from MALT90 data, indicated by white numbers.  The color coding for the plus-sign markers represent the relative match to the LSR velocity of \Pebble:(green indicates a velocity within $\pm 10$ \kms, yellow within $\pm 20$ \kms, and blue all other velocities.  \Pebble\, appears to be associated with a large star-forming complex containing many clumps with similarly large linewidths. }
\label{Figure9}
\end{figure}



\clearpage
\begin{center}
TABLE 1. Molecular Line Properties of \Pebble
\begin{tabular}{ l c c c c }
\hline
 & AGAL337.334$-$0.111 & AGAL337.341$-$0.141 & AGAL337.342$-$0.119 & AGAL337.348$-$0.159  \\
\hline
$V$(N$_2$H$^+$) [\kms]& $-71.4 \pm 0.5$  & $-70.6 \pm 0.4$ & $-72.3 \pm 0.6$ & $-72.5 \pm 0.2$  \\ 
$V$(HNC) [\kms] & $-70.8 \pm 0.3$  & $-71.0 \pm 0.2$ & $-71.8 \pm 0.3$ & $-72.4 \pm 0.2$  \\ 
$V$(HCO$^+$) [\kms]&  $-70.6 \pm 0.2$  & $-70.5 \pm 0.2$ & $-71.9 \pm 0.2$ & $-72.2 \pm 0.1$  \\ 
$\Delta V$(N$_2$H$^+)  [\kms]$ &  $6.1 \pm 0.8$ & $5.9 \pm 0.5$ & $7.6 \pm 1.1$ &  $5.0 \pm 0.3$ \\
$\Delta V$(HNC) [\kms]& $10.6 \pm 0.7$ & $7.6 \pm 0.4$ &  $11.1 \pm 0.7$ & $ 6.7 \pm 0.4$  \\
$\Delta V$(HCO$^+) [\kms]$ &  $15.1 \pm 0.6$ &  $10.6 \pm 0.4$ &  $13.1 \pm 0.5$ &  $8.8 \pm 0.3$ \\
\hline
\end{tabular}
\end{center}


\clearpage
\centerline{TABLE 2. Physical Properties of \Pebble}
\begin{center}
\begin{tabular}{ l c c}
\hline
Parameter & Far Distance &  Near Distance  \\
\hline
Distance     &  11 kpc & 4.7 kpc  \\
Mass           &  27,000 \Msun\, & 5,000 \Msun\, \\
Radius major axis  & 7.7 pc & 3.3 pc \\
Radius minor axis &  1.9 pc & 0.8 pc \\
Average Density & 1400 cm$^{-3}$ & 3300 cm$^{-3}$ \\
Virial mass & $1 \times 10^5$  \Msun\, & $4 \times 10^4$  \Msun\, \\
Virial parameter $\alpha$ & 3.7 & 8.7 \\
Dust temperature & 14 K & 14 K \\
\hline
\end{tabular}
\end{center}


\end{document}